# Biological Hypercomputation: A Concept is Introduced

Carlos Eduardo Maldonado and Nelson Alfonso Gómez-Cruz

Modeling and Simulation Laboratory,
Universidad del Rosario,
Bogotá, Colombia
{carlos.maldonado,nelson.gomez}@urosario.edu.co

**Abstract.** This paper discusses the meaning and scope of biological hypercomputation (BH). The framework here is computational, and from the outset it should be clear that life is not a standard Turing Machine. Living systems hypercompute, but the distinction is made between classical and non-classical hypercomputation. We argue that living processes are non-classical hypercomputation. Yet, BH entails new computational models, for it does not correspond, any longer, to the Turing Machine model of computation. Hence, we introduce BH having a twofold scope, thus: on the one hand, it implies new computational models, while on the other hand we aim at understanding life not by what it *is*, but rather by what it *does*. From a computational point of view, life *hypercomputes*. At the end we sketch out the possibilities, stances and reach of BH. The aim of BH is basically help understanding life from a computational standpoint.

## 1 Introduction

Thanks to the development of computation the sciences of complexity were born [1] but the sciences of complexity have also contributed to the development of computation. The modeling and simulation of such processes and phenomena as non-linearity, adaptation, information and communication, self-organization, and emergence – to name but just a few – made possible to better grasp and see characteristics such as learning, local and global interaction, birth and death, complex networks, for instance. The workings went further to pervade biology at large to-date.

However, theoretical computer science has remained within the framework of classical science, namely the Turing Machine and the (Strong) Church-Turing thesis (CTt) [2]. Certainly, other alternatives have been considered, but engineering on-the-field in general and computer engineering in particular, remain all in all within the mainstream of the CTt and the Turing Machine.

The TM models are no more and no less than algorithmic computation. It is clear that algorithmic computation is closed- the world is shut out -, its resources are finite, and the behavior (modeled) is fixed (conspicuously *top-down*) [3]. As a consequence, the TM does not represent a complete model of computation [4].

The computability problem concerning most of the new paradigms and models inspired by biology have been reduced to the possibility that they can equate the computational power of the universal TM, for they assume, from the outset, that in the Turing model there are limitations about what can and cannot be computed. Among the workings that can be placed in this direction we encounter some models of membrane computing [5]; one-dimensional cellular automata (rule 110) [6] and two-dimension cellular automata (the game of life, Conway) [7], and some DNA computing models [8]. In other words, these can be taken as biological metaphors applied to models of computation.

Proving the Turing-completeness of the new computational paradigms – and particularly in those inspired by biology – has become a matter of routine and yet, according to Syropoulos [9], the computational TM-based paradigm "cannot and should not form the basis for proposing and studying new models of computation inspired by natural processes."

Biological computation, moreover, is a field that has rarely been clearly and deeply worked out, especially within the sciences of complexity. The subject regarding biological computation has been, so far, about the computational models that can be adapted or developed in order to understand biological processes. However, biology is not to be taken exclusively as a metaphor for the aims of computation. On the contrary, the real and defiant task concerns the understanding of what living systems do in order to live. This, we claim leads us onto biological hypercomputation (BH).

BH can be grasped as the explanation of computational processing the biological processes and dynamics ranging from the lowest scales, i.e. genes and bacteria up to the biosphere. Hence, it encompasses both the developmental as well as the evolutionary stances in life. In this paper we introduce the concept of BH throughout four arguments, thus: a) biological systems are not machines, and in which sense they are not; b) evidences in the literature about the plausibility of the concept are brought out. As a consequence, the distinction between classical and non-classical hypercomputation must be made; c) the very concept is introduced and discussed. While we bring out some experimental evidence shown in the literature, we also assess that radically different computational models must be conceived; d) the reach, stances, and possibilities of the concept introduced in this paper are discussed. It is our contention that for the sciences of complexity the concept of BH helps magnificently to better understand what life is and what the biological processes are all about. At the end several conclusions are drawn.

## 2  Life Is Not a (Turing) Machine

Modern science conceives of life –the human body, living systems, and even nature- as a machine very much along the framework of classical mechanics. For modern science a machine is a device clearly defined in anatomical and physiological terms. In the first decades of the 20$^{th}$ Century machines were understood somehow according to the Turing Machine (TM) and hence as a computational device.

Within the computational framework the best characterization of a machine is the TM. Yet, life is *not* a TM, whence new models of computation different from the TM must be developed or introduced. These new models of computation are addressed to understanding what life does in order to live, and not to just to solve computational problems in the traditional sense. As it is well known a TM is an algorithmic device, i.e. machine that can solve P problems in terms of Cook and Karp's computational complexity.

Living systems develop *and* evolve. However neither biological development nor evolution are carried out in algorithmic terms. Evolution is not to be taken in classical Darwinian terms – which, by the way, might indeed be grasped as an algorithmic process – but rather, according to Jablonka and Lamb [10] as the entire process that encompasses the genetic, the epigenetic, the behavioral, and the symbolic levels.

Living systems do compute though not in the sense of the classical TM. A tentative analogue here could be the two sorts of machines Turing himself suggested, namely an automatic machine in respect to the mathematical Hilbert's tenth problem; additionally, a thinking machine, by which he was thinking on logical and philosophical contexts, i.e. a machine that could not be reduced to mere algorithmic processes. There are, indeed, other models of computation that cannot be taken as mere algorithmic proposed by Turing most notably the oracle machine (o-machine), and the choice machine (c-machine). These machines do not compute: they hypercompute, this is, they interact with the environment, and make decisions on their own – provided an external operator.

Computation is for biological systems more than a matter of one tape that runs sequentially. It is matter that entails metabolizing, cellular and homeostatic processes whence health and disease; after all, computation entails for the living systems the edge between life and death. Figure 1 shows the (evolutionary and computational) steps from physical machines to living systems.

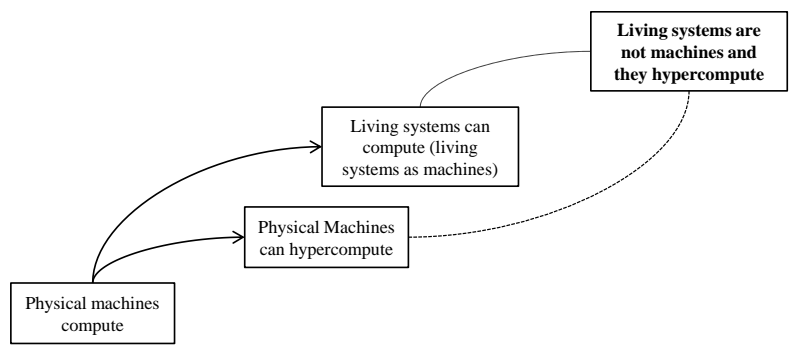

**Fig. 1.** Living systems are not machines and they hypercompute.

However, it should be clearly pointed out that living systems are physical entities and yet cannot be reduced to sheer physics. This is where, according to Jablonka and Lamb [10], the behavioral and symbolic stances are introduced in evolution. The behavioral stance means all capacities to adapt for a given individual whereas the symbolic ones are the outcome of cultural interactions. "Physics" will not be understood as "material" which allows us to encompass "information" as well. Information is a physical entity, indeed. If so, living systems process information as not just material stuff but as knowledge. We will come back to this.

## 3  Hypercomputation: From Classical to Non-Classical

Hypercomputation is concerned with behaviors and phenomena that lie outside the capacities and interests of the TM. However, hypercomputation can be traced back to the workings by Turing [11] – particularly the c-machines and o-machines (see a description in [3]), but remained abandoned or in oblivion for a long time. Isolated works were undertaken in-between, until in 1999 Copeland and Proudfoot [12] coined up the term hypercomputation.

The theory of hypercomputation refers to the theoretical and practical feasibility of computing numbers and functions that the TM is unable to compute [11]. However, hypercomputation not only deals with the computation of numbers and functions that cannot be computed by a Turing Machine, but, more adequately and generically, it looks for finding new ways for solving problems and computing things [9]. Thus, in other words, hypercomputation widens up the framework of what is computable within the context of the CTt while trying to generalize the concept of computation in order to include dynamics, phenomena and processes that were never before taken into account as essentially computable.

In this sense, according to Stepney [13], the models of hypercomputation that seek to solve the halting problem or to compute Turing-uncomputable numbers can be gathered under the generic name of *classical hypercomputation* (CH). Those models of hypercomputation that try to move beyond challenging the traditional notions of computation and the pillars established by the theory of computation are known as *non-classical hypercomputation* (nCH).

CH intends transcend the limitations of the TM, wherein the basis is classical physics, by taking advantage of quantum theory and the general and special theory of relativity [13]. This form of hypercomputation implies the possibility of carrying out super-tasks, that is, for instance, carrying out an infinite number of operations in a finite time – and this would presumably allow for solving in principle Turing's halting problem [9]. In spite of the fact that all models of hypercomputation that go in this direction remain on the theoretical plane, there is nothing, so far, in the known mathematics and physics that impedes the implementation of such hypermachines [14]. Yet, CH remains within the boundaries imposed by the TM model.

As for nCH, it is concerned with widening up the limits of what classically has been understood by "computing". The heuristics Stepney employs for introducing nCH consists in asking several questions, which show the tentative or exploring character of the issues, as well as the innovation of the field. The paradigms pointed out by Stepney as means to reach nCH are related to questions such as the physical embodiment, the use of parallelism as the unique mean to achieve a real time response, the interactive nature of computation (interactive computation), and the plausibility of some kind of emergent hypercomputation.

Within the paradigms brought out by Stepney, *interactive computation* (IC) is the only one that has been formally constituted and broadly developed in the context of hypercomputation [9, 15, 16]. In fact, some of current computing systems such as Internet or sensing and acting robotics [4] are better modeled in terms of interaction than in algorithms. This is, whereas the TM allows us to model closed systems, IC models open systems that interact with environments essentially uncomputable. Moreover, a TM does not know of time, whereas IC knows and works with the arrow of time [15]. Thanks to IC time is openly incorporated within computation. Life is the best example of an interactive system. We strongly believe that IC is the only field within hypercomputation that clearly allows opening the door toward BH, so far.

All in all, hypercomputation does not know of biological computation, and vice versa: biological computation does not know of hypercomputation. Paradoxically they are both stuck with the CTt and the TM's paradigm. To be sure, the openness of computation to complexity produces hypercomputation and, we claim, it is nCH that leads to BH.

## 4   Biological Hypercomputation

Biological computation is a recent research area that studies the information processing that living systems carry out in the natural environment [17]. Within the sciences and disciplines that have contributed to consolidate this field there are natural computing, artificial life, systems biology, synthetic biology, the sciences of computation – most notable the new computational paradigms -, and the sciences of complexity. Biological computation differs conceptually and theoretically from such strands as bio-inspired computation, modeling and computational syntheses of biological processes and the use of biological materials for computational aims. Nonetheless, the frontiers among these strands are currently blurred and hardly identifiable. According to Mitchell [17] "it is only the study of biological computation that asks, if, how, and why living systems can be viewed as fundamentally computational in nature".

In biological computation there are usually three ways to explain biological phenomena as computational concerns. These are: theoretical computing models *(in info)*, descriptions, and experimental observations and simulations. Several examples in the literature about biological computation include bacteria colonies [18]; gene assembly in ciliates [19]; bio-chemical reactions [20]; protein-protein interaction networks, bio-chemical networks, and transport networks [21]; information diffusion in the endocrine system; defense adaptation and coordination in the immune system [22]; information processing in swarm insects [23] and evolution as computation [24].

In order to make the concept of biological computation possible M. Mitchell [17] asks four questions. They concern how is information represented in the system, how information is read and written by the system, how it is processed, and how this information acquires function (or "purpose" or "meaning"), and studies comparisons among various kinds of computation and what life does, computationally speaking. Not far ago, S. Stepney [13] does include the term "biological hypercomputation" and yet she does not develop nor deepen into the concept. Therefore, it is our contention to justify and explain the concept of BH here.

There are, indeed, already a number of theoreticians and researchers who are already working in terms of BH, even though they do not make use of the concept as such. G. Kampis [25] introduces the idea of self-modifying systems, especially in the biomolecular context, and he claims that these systems cannot be expressed or explained in terms of a TM. Moreover, he says that the TM can be applied only to simple and complicated systems. Furthermore, E. Ben-Jacob [18] makes substantial differences between bacterial information processing and the Universal Turing Machine (UTM). Whereas in the UTM the software and hardware are separated, within a bacterial colony the hardware changes according to the input, the information stored, the information processing, and the desirable output. Bacteria can exchange information with each other in the form of hardware (i.e. genetic material).

It should be clear that the Turing machine works on a single scale, sequentially and depending on an external operator. Life, on the other side, processes information locally and is able to synthesize information on higher scales when dealing with given circumstances or solving relevant problems. That is, life does not possess all information needed to live, and that is why she has to synthesize it on the basis of genetic information, memory, and information about and from the environment [18]. Life´s computation is self-organized and emergent.

Emergence is the outcome of self-organization, and by definition a TM is not a self-organized but a top-down controlled machine by an external operator. A TM is a closed system. Living systems, on the contrary are essentially open and incomplete systems. They need matter, energy and information in order to live and they must seek them in the environment. The environment is intrinsically non-computational and non-algorithmic.

The Turing machine is just one particular case of all possible types of computation. According to Dodig [26], computation such as it appears in the natural world is much more natural than what the TM is able to model – or simulate. As a matter of fact, in nature living beings process more than numbers and functions. Along this line, Wolfram [27] speaks of computational irreducibility. That is, there are phenomena which cannot be computationally solved and, therefore, the alternative is to simulate them and watch what happens. Among these kinds of phenomena are the biological systems and processes. In contrast, Mitchell [17] claims that the aim is certainly not just simulate biological systems but understanding the very phenomenon as computation.

It should be explicitly pointed out that the aim of BH does not consist in simulating life but to develop computational models capable of expressing the way life computes. In other words, the issue is not about making (computational) idealizations of life and the living processes but, on the contrary, understanding life and explaining via computation how life computes. We just simply do not know how that computation is carried out though good hints have been provided along the literature [28, 29].

Let us put it straightforwardly: living systems lack an algorithmic solution. Presumably, living processes, for instance, biological, immunological, nervous and others also lack an algorithmic solution. If so, biological hypercomputation is not only possible but necessary. In other words, biological processes – and henceforth living systems – operate on the basis of solving problems, adaptations, learning and self-organization in such a way that life is made possible at the cost of normal algorithmic explanations. Life computes and creates always new functions hardly understandable *ex ante*. Life is concerned with highly complex computational problems such as protein folding, self-repair, interaction with the environment, metabolizing – well, development and evolution. We find ourselves just on the eve of a non-mechanical and non-reductionist understanding and explanation of *what life is and is about by*

*what life does*. To be sure, development and evolution are not about decidability/undecidability.

Science at large is however the story of understanding and making possible, via technological means, non-recursive functions (and more adequately computable nonfunctions). Here is where the sciences of computation and biology play an important role.

As it happens many times in science, scientists and engineers make things without really knowing how things work in nature and why [30]. The examples go from physics to biology, from engineering to astronomy, for instance. To be sure, in the case of living beings, we have come to know that they are not machines - which is not a minor issue! - and that the computation they carry out cannot by any means be reduced nor explained or understood in terms of the CTt. The computational architecture of livings systems is clearly not the Von Neumann architecture of classical computation. Within the framework of non-classical computation other forms of better architectures have been tentatively introduced but always in the context of the TM and the CTt.

BH steps aside of the "classical" idealizations that come to be in each case special cases of more general theories – very much in analogy as Euclidean geometry with respect to non-Euclidean geometries, classical logic vis-à-vis non-classical logics or classical mechanics in relation to the theory of relativity.

The arguments that support BH are:

i) Living beings are not just machines, in any concern; i.e. even though they are physical entities they cannot be understood as mere machines or reduced to physics. New concepts and insights are needed;
ii) Life is the complex phenomenon *par excellence* and it cannot be explained from the inferior or the lower stances. Life, it appears, is to be run and explained as it happens: it is not compressible, in Turing's terms;
iii) The difference between the environment and the living being cannot by any means be traced. Since the environment is an intrinsically open undetermined concept, there is no clear frontier between biotic and a-biotic domains. Even though the environment consists in the atmosphere, the lithosphere, and the hydrosphere, the adaptability and manifold expressions of life do not exist apart from the work living organisms exert on the environment to which they adapt and they help transform. The concept of co-evolution is most suitable here [31];
iv) Interaction – with the environment – is much more expressive than what an algorithm or a TM can be [2]; Interactive computation whether from microorganisms on to human beings [32] or from one species to another to the environment is a matter that clearly opens up the door to BH;
v) Life does not process in terms of classic logical functions – at least as they are known so far-; life is not worried by solving problems in the sense of mathematical functions, whatsoever, either. Other non-classical logics are needed and can be developed and applied here. One of these non-classical logics is paraconsistent logics [33];
vi) In computational terms, for the living systems the difference between the hardware and the software is irrelevant [18]. From a philosophical point of view, dualism can be superseded if not overcome;

vii) For life computation means living; good computing means evolving and bad computing brings living systems to the edge of peril, danger or extinction;

It follows that BH aims both at a rupture and a synthesis wherein the complexity – in the sense of the sciences of complexity – of life comes to the fore. The complexity is exactly the outcome of the asymmetry between the inside and the outside, i.e. between the closed membrane and the autocatalytically closed system of components and the environment [34, 35].

Figure 2 shows the historical development of computational models against a process of complexification. It should be taken as a general view, particularly regarding the time or historical references. It brings forth a comprehensive idea about the origins, ruptures and synthesis wherein biological hypercomputation is born.

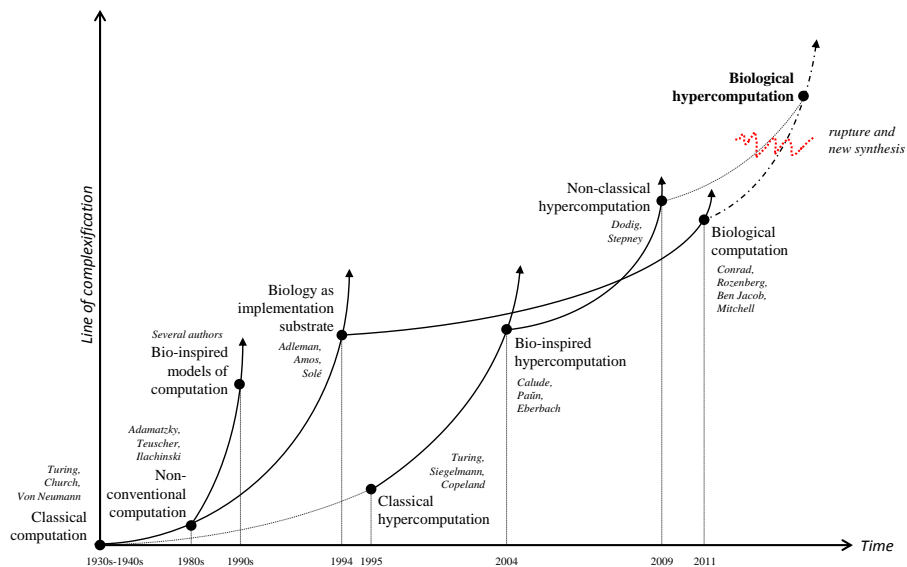

**Fig. 2.** The complexification of computation.

BH opens up a new research field that will transform and at the same time bring closer together biology and computing sciences, philosophy and mathematics, to say the least. The goal consists in understanding and explaining the singularity of life – a singularity that remains within us and yet apart from us, so far. If the universe can be accounted for as a computational stance [36, 37] then the consequence is unavoidable, namely the universe is alive – a conclusion many scientists have reached throughout other ways and roads [34, 38, 39]. This is the final or ultimate framework of biological hypercomputation but the departing gate is to be located in the study of proteins, bacteria, the immunological system, the DNA and RNA strands, on to the organic level and the evolutionary scale.

Let us put it in analytical terms: The standard TM works with one tape on a sequential, i.e. linear scale. Processing is carried out as a black box in the sense that what is relevant is the translation of the Boolean logic into readable universal characters. This is what is called as a "general purpose machine". The processing is based on numerical characters. Processing means literally changing one thing into another, for instance Boolean characters into numbers, figures, letters, and the like. However a TM does never synthesize information. In contrast, life is capable of synthesizing *new* information on different scales. Biological syntheses do transform one thing into another, indeed; however it *produces* brand new information that was not contained any means in the input, as in the TM. BH can adequately be grasped as metabolizing – information.

Life is a network of intertwined scales from bacteria to archaea to eukarya, and vice versa – on to the biosphere. In other words, from genes to cells to organs to systems on to the organism, and the interplay between homeostatic and homeorhetic processes. Life's interaction with the environment is a question of resisting to the physical forces of the environment – homeorhesis. This is what Varela [40] dubs as the autonomy of living systems. However, at the same time once other living systems belong to the environment, the interaction is the interplay of cooperation and competition.

Life is a non-algorithmic phenomenon. An algorithmic system works in three sequential levels, thus: input, processing, output, and they can never be mixed or intertwined. Computation literally happens in the second level, provided the input, and it takes place as a closed system ("black box"). Being essentially open, life never works in clearly stratified or differentiated sequential levels, for energy, matter, and information are constantly entering, being processed and produced in parallel, distributed, cross-linked dynamics. This is where BH happens and this is precisely what is to be understood. Hence, life does synthesize new information and new mechanisms for processing the new information.

All in all, BH is a cross concept that encompasses engineering, scientific – for instance the biological, physical and computational sciences-, logical, and philosophical concerns. It aims at understanding life in computational terms, it shows the need for new computational paradigms, and explains multilevel, parallel, hybrid, and sequential (non-algorithmic) information processes. BH links the microscopic view of life together with the macroscopic view of nature. As such it can be easily linked with one of the non-classical logics, namely quantum logics (besides the call for paraconsistent logics we mentioned above). Thus, BH serves as a unitary though not a universal abstract concept. The concept might be considered as controversial, but it is a bet for a revolution in computer science. A revolution where the main participants are computer scientists working along with biologists, physicists, philosophers, and logicians.

## 5 Possibilities, Stances, and Reach of Biological Hypercomputation

BH entails from one extreme to the other enriching and deepening of the works and understanding of biology and computation. Moreover, biology and computation are, we believe, radically transformed by BH. In an applied sense, the sciences of complexity find in biological hypercomputation its most attractive if not feasible and open way for future research. In the theoretical standpoint, the question remains as to how exactly is BH carried out in every level or scale and how that scale interacts – hypercomputationally! – with the other levels of an organism, a species and the biosphere.

BH does not happen as a black box. First, the processing does not take place after the information is entered (input). Second, the output does not occur as a *sequitur* but is one and the same process as living. As a consequence, thirdly, there are not two stances: the processing unit (PU) and the information, which is by definition external to the PU. There is no gap between the environment and a living system even though there is indeed an asymmetry and that is the key for BH. As it happens, in the TM and the CTt there is symmetry between input and output; that symmetry is linked and "processed" as a black box by the machine. In other words, the information that enters and the information gotten are computationally symmetric. This is exactly what the TM and the Von Neumann architecture are about. The tricky role is played, explained and understood in terms of a black box, or is presumed to be on - tacitly.

More radically, there is no halting state for living beings, for such a state would be simple death. Complexity is about understanding how asymmetry is produced from symmetry, for life is asymmetric – whence, the arrow of time.

On the other side, life is an essentially asymmetric phenomenon that yet arises from symmetry – and breaks it down. This is what is usually called in the literature as dynamic equilibrium. The asymmetry life *is* sheds some new light on the following guess: BH is, correspondingly, an asymmetric processing phenomenon.

The main idea consists in highlighting how living "machines" and processes are not, in any concern, machine or mechanical processes. This claim might sound trivial, but mainstream science still considers – at least *impliciter* – biological dynamics and processes as mechanical. This can be seen throughout the language used, for instance, in neurology where many of the brain processes are still explained in terms of "on" and "off" switches; or in the immunological theory where many processes are understood in terms of "inhibitors" and "excititors". Many other cases can be easily brought out here. In classical terms, there is, though, a certain delay between the information processing and the actions carried out as consequence. The traditional explanation has been in terms of instincts or impulses. The failure, however, is that they have been put out in terms of "on" and "off", which is quite mechanical and hence wrong.

Several philosophical and social implications follow, then. Perhaps one of the most interesting is the promise that BH might transform technology, i.e. engineering [41]. But we can also venture that BH is a multilevel, parallel, and sequential (but non-algorithmic) processing, all at the same time, though not always with the same relevance. Indeed, even though BH remains on the theoretical ground for the time being, there are practical implications to be developed. These implications are, however, the subject of a different text.

## 6 Concluding remarks

This paper introduces the concept of BH. Whereas hypercomputation both opens and demands new computational models, the development of the concept and the reach and scope of the new computational models are issues that remain out of the scope here for reasons of time and space.

BH allows for a dynamic symmetry between computation and biology, for it transforms both in their current and classical sense into a brand new research field. This field is scientific as well as philosophical.

BH promises to open up the concept of computation and what it truly means and entails, and at the same time it makes possible to understand life in a more unified way. Since there is no such a thing as a general theory of biology or of the living systems [16, 32], BH, we believe, can pave the road towards such a theory.

The language about BH is set out in future tense in the literature. This, however, should not be taken in any sense as a mere desideratum or a wild idea or speculation. Scientific, technological and philosophical research is being undertaken toward that future bringing it to the present.

So far, hypercomputation has not been capable of understanding biological processes, and develop a series of tools. In any case, to be sure, there is one more complex arena that arises here, namely how the mind hypercomputes. This question raises questions and possibilities not just for the computation sciences and biology but also for philosophy, epistemology and mathematics if not also for logics. We have aimed at understanding life from a computational point of view; but it should be clear that we do not pretend to reduce life to a computational framework.